# On the nature of the positronic bond

Mohammad Goli[1] and Shant Shahbazian[2]

[1]School of Nano Science, Institute for Research in Fundamental Sciences (IPM), Tehran 19395-5531, Iran, E-mail: m_goli@ipm.ir

[2]Department of Physics, Shahid Beheshti University, G. C., Evin, Tehran, Iran, 19839, P.O. Box 19395-4716. Tel/Fax: 98-21-22431661, E-mail: sh_shahbazian@sbu.ac.ir

**Abstract**

Recently it has been proposed that the positron, the anti-particle analog of the electron, is capable of forming an anti-matter bond in a composite system of two hydride anions and a positron [*Angew. Chem. Int. Ed.* **57**, 8859–8864 (2018)]. In order to dig into the nature of this novel bond the newly developed multi-component quantum theory of atoms in molecules (MC-QTAIM) is applied to this positronic system. The topological analysis reveals that this species is composed of two atoms in molecules, each containing a proton and half of the electronic and the positronic populations. Further analysis elucidates that the electron exchange phenomenon is virtually non-existent between the two atoms and no electronic covalent bond is conceivable in between. On the other hand, it is demonstrated that the positron density enclosed in each atom is capable of stabilizing interactions with the electron density of the neighboring atom. This electrostatic interaction suffices to make the whole system bonded against all dissociation channels. Thus, the positron indeed acts like an anti-matter glue between the two atoms.

# Introduction

The chemistry of exotic species, i.e. molecules containing exotic elementary particles like positron or positively/negatively charged muons, has a venerable history,[1] and in recent years the fields of the muonic,[2,3] and the positronic,[4,5] chemistries became mature subdisciplines of the exotic chemistry. Particularly, understanding positron's interaction with molecules and the concomitant positron-electron annihilation process are of great interest. Beyond the academic curiosity, the annihilation process is the basis of the Positron Emission Tomography as a powerful medicinal imaging technique.[6] Accordingly, a large body of experimental,[7–10] and theoretical and computational,[11–31] studies have been conducted recently on polyatomic and diatomic,[32–39] positronic species in order to trace the sticking site of the positron. These studies reveal that in general the positronic density is very diffuse and is *not* centered between bonds but *behind* the most electronegative atom of the molecule (the cases with two or more atoms with equal electronegativity are more complicated).[20] It is usually perceived that the positron does not participate directly in forming the chemical bonds. Only through the reorganization of electronic structure, which seems to be marginal in general, positron may indirectly affect the bonds between atoms. However, in some very simple positronic species like positronium hydride,[40–43] positronic water,[44,45] and di-positronium,[46–48] there are evidence that the positron is actively participating in bonding interactions. By the way, it is hard to contemplate these species as composed of discernable atoms in molecules, so their classification as molecules is ambivalent and the chemical nature of bonding in these species yet seems to be obscure.

With such background in mind, the recently published paper by Charry, Varella and Reyes (hereafter denoted as CVR) claiming the first unambiguous positronic bond is quite striking.[49] Armed with their newly developed ab initio code, LOWDIN,[50,51] which is capable of dealing with multi-component quantum systems, the authors solved Schrödinger's equation for $\left[e^+, H_2^{-2}\right]$ species. The potential energy surface was derived with sufficient accuracy in order to claim its stability relative to all possible channels of dissociation. Interestingly, by deriving the positron's density and comparing it to the electronic density of some well-known species, e.g. $H_2^+$ and $Li_2^+$, the authors provided some evidence that the positron is the main bonding agent acting as a *glue* between the two hydride ions. Based on these findings, it seems reasonable to symbolize this species as $\left[H^-, e^+, H^-\right]$. The fact that the positron's density is maximum between the two hydrides was interpreted by CVR as a manifestation of a one-positron covalent bond. Our aim in this communication is to verify detailed nature of the proposed positronic bond.

**Results and discussion**

In order to consider the positronic bond we employ the recently developed multi-component quantum theory of atoms in molecules (MC-QTAIM),[52–59] which is an extended version of the QTAIM proposed originally by Bader and coworkers.[60] The MC-QTAIM is the only available chemical theory specially designed to deal with the bonding analysis of the exotic species. The MC-QTAIM analysis is done taking into the number density and property densities of all types of quantum particles (not just those of electrons'). Using ab initio derived multi-component wavefunction of an exotic species, through a well-defined and unique machinery, which is system-independent and automated, the MC-QTAIM analysis derives the AIM and their properties. These properties may then be used to access the bonding modes of the AIM in the exotic molecule as has been previously done in the case of the positronic,[61–63] and the muonic,[64–68] species. Particularly, the previous MC-QTAIM analysis of the positronic species revealed that the positron is not capable of accumulating enough electrons around itself to form an independent atomic basin.[62] In all the considered species the positron retains in the basin of the most electronegative atom except from the case of cyanide anion, $CN^-$, where the positron's population was almost evenly distributed in both atomic basins.[62] In present analysis we employ the MC-QTAIM trying to reveal the detailed nature of the AIM structure as well as the bonding mode of $\left[H^-, e^+, H^-\right]$.

Let us first very briefly discuss the energetics and stability of $\left[H^-, e^+, H^-\right]$ where CVR considered this species at various multi-component ab initio levels. The used computational levels were the MC-HF, the MC-MP2 and the MC-CI, combined with the standard correlation consistent basis sets for *both* electrons and the positron (hereafter the first basis set in the parenthesis is for electrons and the second one is for the positron).[49,50] At the highest ab initio levels, i.e. MC-CISDTQ/(aug-cc-pVDZ/aug-cc-pVDZ) and MC-CISDTQ/(aug-cc-pVTZ/aug-cc-pVTZ), the computed binding energies (BEs) relative to dissociation to the positronium hydride and the hydride ion, were ~55 and ~66 kJ.mol$^{-1}$, respectively.[49] We employed the same basis sets and the NEO computer code,[69,70] with some modifications, to compute BEs at the MC-HF level. At fixed 3.2 Å inter-nuclear distance, derived as the equilibrium point at the MC-CISDTQ level,[49] the BEs are ~36 (aug-cc-pVDZ/aug-cc-pVDZ) and ~49 (aug-cc-pVTZ/aug-cc-pVTZ) kJ.mol$^{-1}$ (for details see Tables S1 in the supporting information). These are not very accurate values compared to those computed at the highest correlated level. But, clearly demonstrate that even at the MC-HF level the system is bound and the computed BEs recover more than 65% (aug-cc-pVDZ/aug-cc-pVDZ) and 75% (aug-cc-pVTZ/aug-cc-pVTZ) of the exact BEs. This observation justifies employing the MC-HF wavefunction for further bonding analysis, as also used by CVR,[49] since the energetic origins of the binding must be present also at this computational level.

The whole MC-QTAIM analysis was done using MC-HF/(aug-cc-pVTZ/aug-cc-pVTZ) wavefunction produced during the ab initio calculations. At first, the electronic, the positronic and the Gamma densities were produced.[52] The Gamma density for the positronic systems is simply the sum of the electronic and the positronic densities,[61–63] and is the basic scalar field used for the topological analysis and deducing the AIM boundaries within the context of the MC-QTAIM analysis. Figure 1 depicts these densities and the minus of the Laplacian of the positronic density, which acts like a magnifying glass, revealing the concentration and depletion of the positronic density. In line with the results reported by CVR the positronic density is concentrated in the middle of the two nuclei and depleted around each nucleus. The topological analysis of the Gamma density reveals two (3, -3) critical points (CPs) at the nuclei and a (3, -1) CP at the middle of the two nuclei. This topological structure, depicted in panel (e) of the figure, is the typical of diatomic molecules.[60] The concentration of the positronic density is not enough to shape a local maximum in the Gamma density at the middle of the two nuclei. Thus, the positron is not capable of forming its own atomic basin in this molecule (even adding a number of extra basis functions to the positronic basis set at the midpoint between the nuclei did not alter this pattern). As discussed recently in details,[71–73] not the analysis of the topological structure nor the amounts of various property densities at (3, -1) CPs are safe grounds to decipher the nature of AIM interactions.

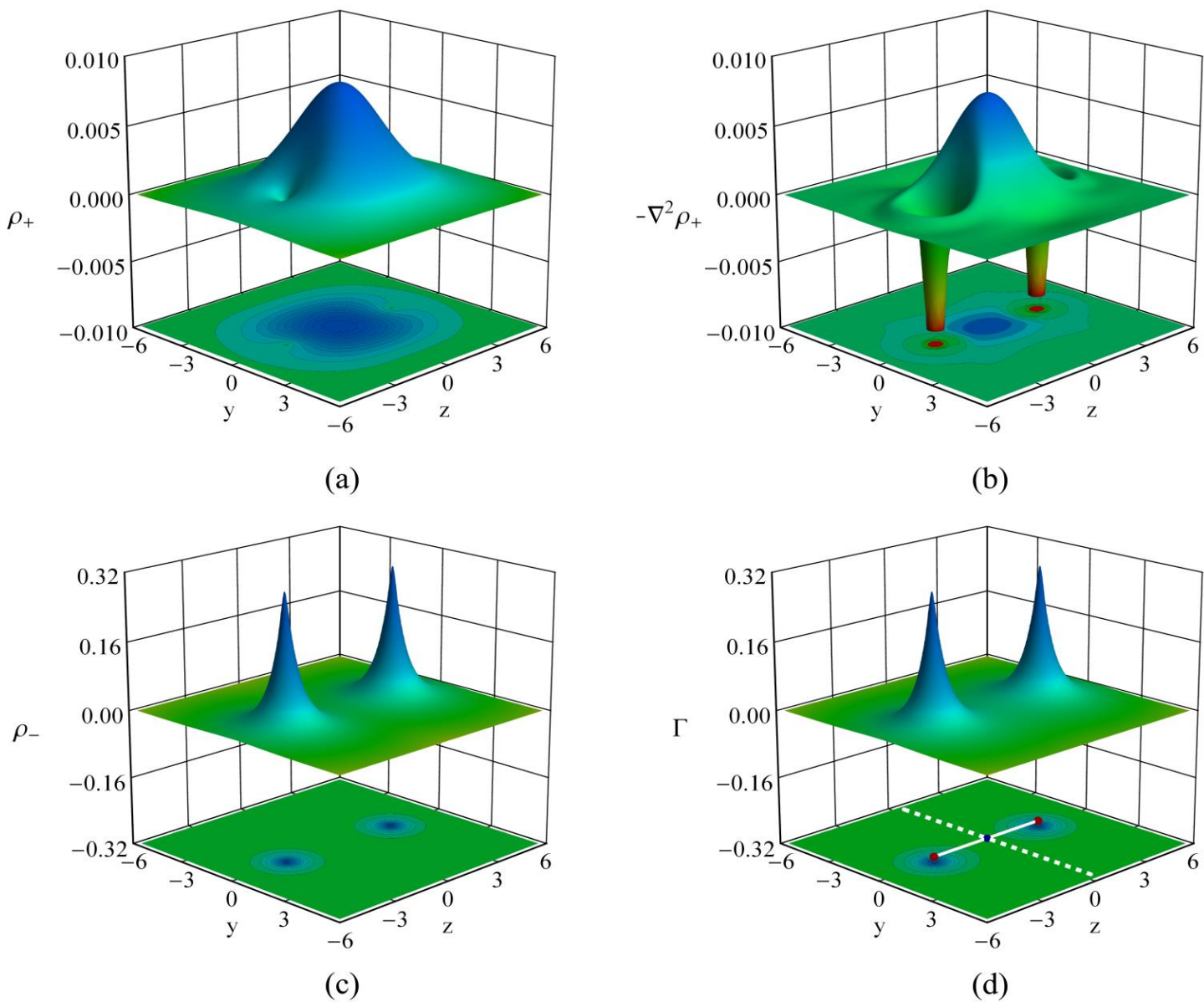

Figure 1. The relief maps of (a) the positronic density, (b) the Laplacian of the positronic density, (c) the electronic density, and (d) the Gamma density of $[H^-, e^+, H^-]$ computed at the MC-HF/(aug-cc-pVTZ/aug-cc-pVTZ) level (the qualitative aspects of these plots are independent from the used basis sets). The red and black spheres in panel (d) are (3, -3) and (3, -1) CPs of the Gamma density, respectively, while the white thread is duo of gradient paths connecting (3, -3) CPs to (3, -1) CP. The dashed white line is the intersection of the boundary of the AIM and the yz plane (in 3D, the boundary is a flat plane perpendicular to the yz plane) revealing the fact that the 3D space is equally divided between the two AIM. All quantities are offered in atomic units.

Instead, the atomic properties must be used to dig into the nature of the interactions.[74] Particularly, in the case of the exotic species there is no firm reason to believe that the usual chemical intuition, which is shaped based on studying purely electronic systems, is trustworthy.

Currently, the best available indices to probe the nature of bonding between the AIM are the electronic delocalization index (e-DI),[75–80] and the energy partitioning based on the Interacting Quantum Atoms (IQA) methodology.[81–85] The extended version of the delocalization indices and certain elements of the extended IQA, introduced within the context of the MC-QTAIM, have been reported elsewhere.[57,58] A detailed account on the extended IQA for the positronic systems is given in the appendix of the paper and only the energy terms relevant to our study is considered herein (the full numerical results of the electronic population and the extended IQA analyzes are given in Tables S2 and S3 in the supporting information). As is evident from Figure 1, the two AIM are completely equivalent and each contains exactly half of the populations of electrons, $N_e(H_1) = N_e(H_2) = 2$, and positron, $N_p(H_1) = N_p(H_2) = 0.5$. Thus, no charge transfer mechanism of bonding is conceivable in this system. The computed extended e-DI for $[H^-, e^+, H^-]$ is also small, ~0.14, and is virtually the same with that computed for $H_2^{-2}$, ~0.15, at HF/aug-cc-pVTZ level at the same inter-nuclear distance. It is a well-established fact that e-DI is a measure of the inter-basin electron number fluctuations, revealing the covalency of the bond (in ionic systems or in the case of the inter-molecular interactions, lacking covalency, its value usually diminishes below 0.2).[78,86] The computed inter-basin electronic exchange energy, $V_{ee}^{xc}(H_1, H_2) \approx -38 \, kJ.mol^{-1}$ ($V_{ee}^{ex}(H_1, H_2) \approx -32 \ kJ.mol^{-1}$ in the case of $H_2^{-2}$), which is an energetic measure of the covalency,[87,88] is also appreciably small compared to those derived from the IQA analysis of classical covalent systems.[84,88,89] Since $H_2^{-2}$ is an unbound system without a discernable covalent bond,[49] it seems reasonable to conclude that the origin of bonding in $\left[H^-, e^+, H^-\right]$ cannot be traced in the electronic exchange phenomenon. To proceed further, the remaining energy terms of the extended IQA method which includes all classical coulombic interactions are considered. These are categorized into two main groups namely, purely electronic group of terms, $V_{elec}^{cl}$, and the group of terms exclusive to the positronic systems, $V_p^{cl}$. Although the interaction of electron density of each basin and the nucleus of the neighboring basin is stabilizing, $V_{en}(H_1, H_2) = V_{en}(H_2, H_1) \approx -851 \ kJ.mol^{-1}$, overall, the purely electronic group is destabilizing, $V_{elec}^{cl}(H_1, H_2) \approx 401 \ kJ.mol^{-1}$ (this is also the case for $H_2^{-2}, V_{elec}^{cl}(H_1, H_2) \approx 311 \ kJ.mol^{-1}$). Thus, the purely electronic group of the classical coulombic interactions is not responsible for the bonding. The only remaining stabilizing classical term is the interaction of electronic density of each basin and the positronic density of the neighboring basin, $V_{ep}(H_1, H_2) = V_{ep}(H_2, H_1) \approx -482 \ kJ.mol^{-1}$. This term suffices for an overall stabilization of the positronic classical group of terms, $V_p^{cl}(H_1, H_2) \approx -471 \ kJ.mol^{-1}$, in contrast to the destabilizing interaction of the positronic density of each basin with the nucleus of the neighboring basin, $V_{pn}(H_1, H_2) = V_{pn}(H_2, H_1) \approx 246 \ kJ.mol^{-1}$. Taking into account all these terms together, the total interaction energy between the two basins, $E_{\text{inter}}(H_1,H_2) = V_{elec}^{cl}(H_1,H_2) + V_p^{cl}(H_1,H_2) + V_{ee}^{xc}(H_1,H_2) \approx -109 \ kJ.mol^{-1}$, is stabilizing.

## Conclusion

Clearly, the dominant cause of bonding in $\left[H^-, e^+, H^-\right]$ is the positron density in the middle of the two nuclei that upon interaction with the electron density acts as a *positronic glue* to transform the original unbound $H_2^{-2}$ into a bonded system. One may ask the delicate question whether this positronic glue is an anti-matter covalent bond, as proposed by CVR, or it is a new type of bond foreign to the purely electronic systems. In a single-positron system there is no positronic exchange energy in the extended IQA analysis as the hallmark of the covalent bond. Instead, the analysis conceives the positronic glue more like a static charge distribution shaping a bond grounded on classical electrostatics although this bond is quite distinct from a classic ionic bond which is linked to charge transfer mechanism. By the way, a recent study by Nascimento and coworkers,[90] trying to trace the interference effects and concomitant energy analysis demonstrates that the basic mechanism behind the formation of one- and two-electron covalent bonds is not much different. In principle, this may also apply to the positronic bond as well though no similar analysis has been developed yet for the positronic systems. Thus, we believe at current state of affairs it is premature to claim a firm answer the above posed question and more studies needed to be done. Taking the fact that the bonding analysis of the exotic system is an undeveloped field of research, the positronic bond is a real challenge to traditional chemical theories of bonding while it is also an opportunity for practitioners of bonding theories to further develop and extrapolate these methodologies to deal with the realm of the exotic species.

**Appendix- The extended IQA analysis**

The IQA partitioning of the total energy of a molecule containing $M$ atomic basins, $\Omega$, in real space is as follows:[82]

$$E_{mol} = \sum_{i}^{M} E_{intra}(\Omega_i) + \sum_{i}^{M}\sum_{j>i}^{M} E_{inter}(\Omega_i, \Omega_j)$$

$$E_{intra}(\Omega_i) = T_e(\Omega_i) + V_{en}(\Omega_i) + V_{ee}^{cl}(\Omega_i) + V_{ee}^{xc}(\Omega_i)$$

$$E_{inter}(\Omega_i, \Omega_j) = V_{en}(\Omega_i, \Omega_j) + V_{en}(\Omega_j, \Omega_i) + V_{ee}^{cl}(\Omega_i, \Omega_j) + V_{ee}^{xc}(\Omega_i, \Omega_j) + V_{nn}(\Omega_i, \Omega_j)$$

(1)

In these equations $(e, n)$ symbols stand for electronic and nuclear, respectively, while $(T, V)$ symbols stand for kinetic and various potential energy terms. The detailed form of each term in the intra-basin group is given below:[82]

$$T_e(\Omega_i) = (-\hbar^2/2m_e) \int_{\Omega_i} d\vec{r}\, \nabla^2 \rho^{(1)}(\vec{r}, \vec{r}')\big|_{\vec{r}'=\vec{r}}, \quad V_{en}(\Omega_i) = -Z_i \int_{\Omega_i} d\vec{r}\, \frac{\rho(\vec{r})}{|\vec{r} - \vec{R}_i|}$$

$$V_{ee}^{cl}(\Omega_i) = (1/2) \int_{\Omega_i} d\vec{r}_1 \int_{\Omega_i} d\vec{r}_2 \frac{\rho(\vec{r}_1)\rho(\vec{r}_2)}{|\vec{r}_1 - \vec{r}_2|}, \quad V_{ee}^{xc}(\Omega_i) = (1/2) \int_{\Omega_i} d\vec{r}_1 \int_{\Omega_i} d\vec{r}_2 \frac{\rho_{xc}(\vec{r}_1, \vec{r}_2)}{|\vec{r}_1 - \vec{r}_2|}$$

(2)

In these equations $(N_e, m_e, Z_i)$ stand for the number of electrons, the mass of electron and the atomic number of the nucleus in $i-th$ basin while the electron density, reduced spinless first and second-order density matrices, and the partitioning of the latter to the classic and exchange-correlation parts, are introduced as follows:

$$\rho(\vec{r}) = N_e \sum_{spins} \int d\vec{r}_2 ... \int d\vec{r}_{N_e} \Psi^*(\vec{r}, \vec{r}_2, ..., \vec{r}_{N_e}, \{spins\}) \Psi(\vec{r}, \vec{r}_2, ..., \vec{r}_{N_e}, \{spins\})$$

$$\rho^{(1)}(\vec{r},\vec{r}\,') = N_e \sum_{spins} \int d\vec{r}_2 ... \int d\vec{r}_{N_e} \Psi^*(\vec{r}\,',\vec{r}_2,...,\vec{r}_{N_e},\{spins\}) \Psi(\vec{r},\vec{r}_2,...,\vec{r}_{N_e},\{spins\})$$

$$\rho^{(2)}(\vec{r}_1,\vec{r}_2) = N_e(N_e-1) \sum_{spins} \int d\vec{r}_3 ... \int d\vec{r}_{N_e} \Psi^*(\vec{r}_1,\vec{r}_2,\vec{r}_3,...,\vec{r}_{N_e},\{spins\}) \Psi(\vec{r}_1,\vec{r}_2,\vec{r}_3,...,\vec{r}_{N_e},\{spins\})$$

$$\rho^{(2)}(\vec{r}_1,\vec{r}_2) = \rho(\vec{r}_1)\rho(\vec{r}_2) + \rho_{xc}(\vec{r}_1,\vec{r}_2) \tag{3}$$

For the inter-basin group, the detailed forms are as follows:[82]

$$V_{en}(\Omega_i,\Omega_j) = -Z_i \int_{\Omega_j} d\vec{r} \frac{\rho(\vec{r})}{|\vec{r}-\vec{R}_i|}, \quad V_{en}(\Omega_j,\Omega_i) = -Z_j \int_{\Omega_i} d\vec{r} \frac{\rho(\vec{r})}{|\vec{r}-\vec{R}_j|}$$

$$V_{ee}^{cl}(\Omega_i,\Omega_j) = \int_{\Omega_i} d\vec{r}_1 \int_{\Omega_j} d\vec{r}_2 \frac{\rho(\vec{r}_1)\rho(\vec{r}_2)}{|\vec{r}_1-\vec{r}_2|}, \quad V_{ee}^{xc}(\Omega_i,\Omega_j) = \int_{\Omega_i} d\vec{r}_1 \int_{\Omega_j} d\vec{r}_2 \frac{\rho_{xc}(\vec{r}_1,\vec{r}_2)}{|\vec{r}_1-\vec{r}_2|}$$

$$V_{nn}(\Omega_i,\Omega_j) = \frac{Z_i Z_j}{|\vec{R}_i-\vec{R}_j|} \tag{4}$$

The origin of bonding is traced in this group and $V_{ee}^{xc}(\Omega_i,\Omega_j)$ is usually attributed to the covalent bonding originating from electron exchange phenomenon while the remaining terms, $V_{en}(\Omega_i,\Omega_j)$, $V_{en}(\Omega_j,\Omega_i)$, $V_{ee}^{cl}(\Omega_i,\Omega_j)$, $V_{nn}(\Omega_i,\Omega_j)$, are usually conceived collectively as classical coulombic interactions and attributed to the ionic bonding.[87,88] Thus, the inter-basin interaction is rewritten as follows:

$$V_{total}^{cl}(\Omega_i,\Omega_j) = V_{en}(\Omega_i,\Omega_j) + V_{en}(\Omega_j,\Omega_i) + V_{ee}^{cl}(\Omega_i,\Omega_j) + V_{nn}(\Omega_i,\Omega_j)$$

$$E_{inter}(\Omega_i,\Omega_j) = V_{total}^{cl}(\Omega_i,\Omega_j) + V_{ee}^{xc}(\Omega_i,\Omega_j) \tag{5}$$

After adding a positron to the molecule and assuming that the number of basins is not varied in this process, the extended IQA partitioning of the total energy of the positronic molecule is as follows:

$$E_{mol} = \sum_{i}^{M} E_{intra}(\Omega_i) + \sum_{i}^{M} \sum_{j>i}^{M} E_{inter}(\Omega_i, \Omega_j)$$

$$E_{intra}(\Omega_i) = T_e(\Omega_i) + T_p(\Omega_i) + V_{en}(\Omega_i) + V_{ee}^{cl}(\Omega_i) + V_{ee}^{xc}(\Omega_i) + V_{ep}(\Omega_i) + V_{pn}(\Omega_i)$$

$$E_{inter}(\Omega_i, \Omega_j) = V_{en}(\Omega_i, \Omega_j) + V_{en}(\Omega_j, \Omega_i) + V_{ee}^{cl}(\Omega_i, \Omega_j) + V_{ee}^{xc}(\Omega_i, \Omega_j) + V_{nn}(\Omega_i, \Omega_j)$$

$$V_{ep}(\Omega_i, \Omega_j) + V_{ep}(\Omega_j, \Omega_i) + V_{pn}(\Omega_i, \Omega_j) + V_{pn}(\Omega_j, \Omega_i) \quad (6)$$

Although the same symbols used for electronic terms in equations (1) are also used for the extended version, it must be stressed that all these electronic terms are reintroduced based on the wavefunction of the positronic system. The intra-basin terms are as follows:

$$T_e(\Omega_i) = \left(-\hbar^2/2m_e\right) \int_{\Omega_i} d\vec{r}_e \, \nabla^2 \rho_e^{(1)}(\vec{r}_e, \vec{r}_e')\big|_{\vec{r}_e' = \vec{r}_e}, \quad V_{en}(\Omega_i) = -Z_i \int_{\Omega_i} d\vec{r}_e \frac{\rho_e(\vec{r}_e)}{|\vec{r}_e - \vec{R}_i|}$$

$$V_{ee}^{cl}(\Omega_i) = (1/2) \int_{\Omega_i} d\vec{r}_1 \int_{\Omega_i} d\vec{r}_2 \frac{\rho_e(\vec{r}_1)\rho_e(\vec{r}_2)}{|\vec{r}_1 - \vec{r}_2|}, \quad V_{ee}^{xc}(\Omega_i) = (1/2) \int_{\Omega_i} d\vec{r}_1 \int_{\Omega_i} d\vec{r}_2 \frac{\rho_{xc}(\vec{r}_1, \vec{r}_2)}{|\vec{r}_1 - \vec{r}_2|}$$

$$T_p(\Omega_i) = \left(-\hbar^2/2m_p\right) \int_{\Omega_i} d\vec{r}_p \, \nabla^2 \rho_p^{(1)}(\vec{r}_p, \vec{r}_p')\big|_{\vec{r}_p' = \vec{r}_p}$$

$$V_{ep}(\Omega_i) = -\int_{\Omega_i} d\vec{r}_e \int_{\Omega_i} d\vec{r}_p \frac{\rho_e(\vec{r}_e)\rho_p(\vec{r}_p)}{|\vec{r}_e - \vec{r}_p|}, \quad V_{pn}(\Omega_i) = Z_i \int_{\Omega_i} d\vec{r}_p \frac{\rho_p(\vec{r}_p)}{|\vec{r}_p - \vec{R}_i|} \quad (7)$$

Let us stress that the masses of electron and positron are the same, $m_p = m_e$, but the charges have opposite signs which manifests itself in the opposite signs of the electron-nucleus, $V_{en}(\Omega_i)$, and positron-nucleus, $V_{pn}(\Omega_i)$, terms. The electron density, the positron density, the reduced electronic and positronic spinless first-order density matrices, the second-order electronic density matrices, and the partitioning of the latter to the classic and exchange-correlation parts, are introduced as follows:[57,58]

$$\rho_e(\vec{r}_e) = N_e \sum_{spins} \int d\vec{r}_p \int d\vec{r}_2 ... \int d\vec{r}_{N_e} \Psi^*(\vec{r}_e, \vec{r}_p, \vec{r}_2, ..., \vec{r}_{N_e}, \{spins\}) \Psi(\vec{r}_e, \vec{r}_p, \vec{r}_2, ..., \vec{r}_{N_e}, \{spins\})$$

$$\rho_p(\vec{r}_p) = \sum_{spin} \int d\vec{r}_1 ... \int d\vec{r}_{N_e} \Psi^*(\vec{r}_p, \vec{r}_1, \vec{r}_2, ..., \vec{r}_{N_e}, \{spins\}) \Psi(\vec{r}_p, \vec{r}_1, \vec{r}_2, ..., \vec{r}_{N_e}, \{spins\})$$

$$\rho_e^{(1)}(\vec{r}_e, \vec{r}_e') = N_e \sum_{spins} \int d\vec{r}_p \int d\vec{r}_2 ... \int d\vec{r}_{N_e} \Psi^*(\vec{r}_e', \vec{r}_p, \vec{r}_2, ..., \vec{r}_{N_e}, \{spins\}) \Psi(\vec{r}_e, \vec{r}_p, \vec{r}_2, ..., \vec{r}_{N_e}, \{spins\})$$

$$\rho_p^{(1)}(\vec{r}_p, \vec{r}_p') = \sum_{spin} \int d\vec{r}_1 ... \int d\vec{r}_{N_e} \Psi^*(\vec{r}_p', \vec{r}_1, ..., \vec{r}_{N_e}, \{spins\}) \Psi(\vec{r}_p, \vec{r}_1, ..., \vec{r}_{N_e}, \{spins\})$$

$$\rho^{(2)}(\vec{r}_1, \vec{r}_2) = N_e(N_e - 1) \sum_{spins} \int d\vec{r}_p \int d\vec{r}_3 ... \int d\vec{r}_{N_e} \Psi^*(\vec{r}_p, \vec{r}_1, \vec{r}_2, \vec{r}_3, ..., \vec{r}_{N_e}, \{spins\}) \Psi(\vec{r}_p, \vec{r}_1, \vec{r}_2, \vec{r}_3, ..., \vec{r}_{N_e}, \{spins\})$$

$$\rho^{(2)}(\vec{r}_1, \vec{r}_2) = \rho(\vec{r}_1)\rho(\vec{r}_2) + \rho_{xc}(\vec{r}_1, \vec{r}_2) \qquad (8)$$

The inter-basin terms are introduced as follows:

$$V_{en}(\Omega_i, \Omega_j) = -Z_i \int_{\Omega_j} d\vec{r}_e \frac{\rho_e(\vec{r}_e)}{|\vec{r}_e - \vec{R}_i|}, \qquad V_{en}(\Omega_j, \Omega_i) = -Z_j \int_{\Omega_i} d\vec{r}_e \frac{\rho(\vec{r}_e)}{|\vec{r}_e - \vec{R}_j|}$$

$$V_{ee}^{cl}(\Omega_i, \Omega_j) = \int_{\Omega_i} d\vec{r}_1 \int_{\Omega_j} d\vec{r}_2 \frac{\rho_e(\vec{r}_1)\rho_e(\vec{r}_2)}{|\vec{r}_1 - \vec{r}_2|}, \quad V_{ee}^{xc}(\Omega_i, \Omega_j) = \int_{\Omega_i} d\vec{r}_1 \int_{\Omega_j} d\vec{r}_2 \frac{\rho_{xc}(\vec{r}_1, \vec{r}_2)}{|\vec{r}_1 - \vec{r}_2|}$$

$$V_{nn}(\Omega_i, \Omega_j) = \frac{Z_i Z_j}{|\vec{R}_i - \vec{R}_j|}, \qquad V_{ep}(\Omega_i, \Omega_j) = -\int_{\Omega_i} d\vec{r}_e \int_{\Omega_j} d\vec{r}_p \frac{\rho_e(\vec{r}_e)\rho_p(\vec{r}_p)}{|\vec{r}_e - \vec{r}_p|}$$

$$V_{ep}(\Omega_j, \Omega_i) = -\int_{\Omega_j} d\vec{r}_e \int_{\Omega_i} d\vec{r}_p \frac{\rho_e(\vec{r}_e)\rho_p(\vec{r}_p)}{|\vec{r}_e - \vec{r}_p|}, \qquad V_{pn}(\Omega_i, \Omega_j) = Z_i \int_{\Omega_j} d\vec{r}_p \frac{\rho_p(\vec{r}_p)}{|\vec{r}_p - \vec{R}_i|}$$

$$V_{pn}(\Omega_j, \Omega_i) = Z_j \int_{\Omega_i} d\vec{r}_p \frac{\rho_p(\vec{r}_p)}{|\vec{r}_p - \vec{R}_j|} \qquad (9)$$

Once again only $V_{ee}^{xc}(\Omega_i,\Omega_j)$ is attributed to the covalent bonding and the remaining terms are classical coulombic interactions. For further analysis, the latter terms may also be divided into two categorizes; the first group are those terms with counterparts in the IQA analysis of the purely electronic systems, $V_{elec}^{cl}$, and second group are those exclusive to the positronic systems, $V_p^{cl}$. Accordingly, the final equations are as follows:

$$V_{elec}^{cl}(\Omega_i,\Omega_j) = V_{en}(\Omega_i,\Omega_j) + V_{en}(\Omega_j,\Omega_i) + V_{ee}^{cl}(\Omega_i,\Omega_j) + V_{nn}(\Omega_i,\Omega_j)$$

$$V_p^{cl}(\Omega_i,\Omega_j) = V_{ep}(\Omega_i,\Omega_j) + V_{ep}(\Omega_j,\Omega_i) + V_{pn}(\Omega_i,\Omega_j) + V_{pn}(\Omega_j,\Omega_i)$$

$$E_{inter}(\Omega_i,\Omega_j) = V_{elec}^{cl}(\Omega_i,\Omega_j) + V_p^{cl}(\Omega_i,\Omega_j) + V_{ee}^{xc}(\Omega_i,\Omega_j) \quad (10)$$

In practice, the origin of stabilizing interactions in the classic electronic terms is the electron-nucleus interaction, $V_{en}$, while in the group of the positronic terms, the electron-positron interaction term, $V_{ep}$, is responsible for stabilizing interactions. All the remaining terms are destabilizing by their nature.

**TOC**

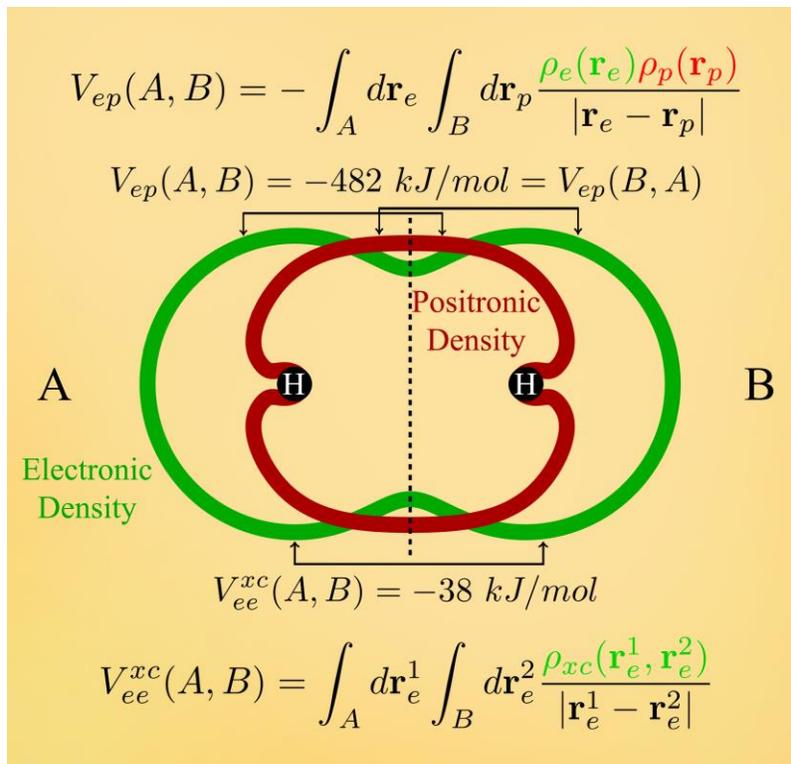

**Incredible anti-matter bond**! The bonding analysis of a composite system consists of two hydride anions and a positron, the anti-matter analog of electron, through the newly developed multi-component Quantum Theory of Atoms in Molecules (MC-QTAIM), reveals that the positron is responsible for a novel electrostatic bond which glues the two hydrides, preventing fragmentation of the composite system.

# Supporting Information

# On the nature of the positronic bond

Mohmmad Goli[1] and Shant Shahbazian[2]

[1]School of Nano Science, Institute for Research in Fundamental Sciences (IPM), Tehran 19395-5531, Iran, Email: m_goli@ipm.ir

[2]Department of Physics, Shahid Beheshti University, G. C., Evin, Tehran, Iran, 19839, P.O. Box 19395-4716. Tel/Fax: 98-21-22431661, E-mail: sh_shahbazian@sbu.ac.ir

# Table of contents

Table S1- The computed energetics (without the counterpoise correction) at the MC-Hartree-Fock level for the target species as well the positronium hydride (PsH) and the hydride ion (H⁻) employing (auc-cc-pVDZ/aug-cc-pVDZ) and (aug-cc-pVTZ/aug-cc-pVTZ) basis sets. In the first column "e", "p" and "n" are abbreviations for the electronic, the positronic and the nuclear, respectively, while "E" and "V/T" are the total energy and the virial ratio, respectively. All values are in Hartrees.

|  | [H⁻,e⁺,H⁻]* | | PsH | | H⁻ | |
| --- | --- | --- | --- | --- | --- | --- |
|  | DZ/DZ | TZ/TZ | DZ/DZ | TZ/TZ | DZ/DZ | TZ/TZ |
| *kinetic* | | | | | | |
| e | 1.12371 | 1.13828 | 0.59293 | 0.59837 | 0.48201 | 0.48755 |
| p | 0.07758 | 0.08567 | 0.06416 | 0.06644 | - | - |
| *potential* | | | | | | |
| ee | 1.52774 | 1.53767 | 0.46808 | 0.46662 | 0.39687 | 0.39641 |
| pe | -0.98426 | -1.01226 | -0.57860 | -0.58315 | - | - |
| en | -3.61992 | -3.64141 | -1.53573 | -1.54106 | -1.36565 | -1.37161 |
| pn | 0.54534 | 0.55440 | 0.32537 | 0.32670 | - | - |
| nn | 0.16537 | 0.16537 | - | - | - | - |
| E | -1.16444 | -1.17229 | -0.66380 | -0.66607 | -0.48678 | -0.48765 |
| V/T | -1.96932 | -1.95779 | -2.01021 | -2.00189 | -2.00990 | -2.00020 |

* The internuclear distance is fixed at 3.2 Angstroms.

Table S2- The computed electronic population (Ne), the extended electronic localization index (e-LI) and the extended electronic delocalization index (e-DI) for [H$^-$,e$^+$,H$^-$] system at HF/(aug-cc-pVDZ/aug-cc-pVDZ) and HF/(aug-cc-pVTZ/aug-cc-pVTZ) levels.

| TZ/TZ AIM | Ne | e-LI | e-DI |
|---|---|---|---|
| H | 2.000 | 1.932 | 0.136 |
| Total | 4.000 | 3.864 | |

| DZ/DZ AIM | Ne | e-LI | e-DI |
|---|---|---|---|
| H | 2.000 | 1.933 | 0.134 |
| Total | 4.000 | 3.866 | |

Table S3- The extended-IQA terms computed for [H⁻,e⁺,H⁻] system at HF/(aug-cc-pVDZ/aug-cc-pVDZ) and HF/(aug-cc-pVTZ/aug-cc-pVTZ) levels. All values are in Hartrees.

| **TZ/TZ** | $E_{intra}$ | $T_e$ | $V_{ne}$ | $V_{ee}^{cl}$ | $V_{xc}$ | | |
|---|---|---|---|---|---|---|---|
| | | 0.569 | -1.497 | 0.899 | -0.441 | | |
| | | $T_p$ | $V_{pn}$ | $V_{ep}$ | | | |
| | | 0.043 | 0.183 | -0.322 | | | |
| | $E_{inter}$ | $V_{en}(H_1,H_2)$ | $V_{en}(H_2,H_1)$ | $V_{nn}$ | $V_{ee}^{cl}$ | $V_{elec}^{cl}$ | $V_{xc}$ |
| | | -0.324 | -0.324 | 0.165 | 0.636 | 0.153 | -0.015 |
| | | $V_{ep}(H_1,H_2)$ | $V_{ep}(H_2,H_1)$ | $V_{np}(H_1,H_2)$ | $V_{np}(H_2,H_1)$ | $V_p^{cl}$ | |
| | -0.042 | -0.184 | -0.184 | 0.094 | 0.094 | -0.180 | |
| **DZ/DZ** | $E_{intra}$ | $T_e$ | $V_{ne}$ | $V_{ee}^{cl}$ | $V_{xc}$ | | |
| | | 0.562 | -1.488 | 0.897 | -0.440 | | |
| | | $T_p$ | $V_{pn}$ | $V_{ep}$ | | | |
| | | 0.039 | 0.184 | -0.319 | | | |
| | $E_{inter}$ | $V_{en}(H_1,H_2)$ | $V_{en}(H_2,H_1)$ | $V_{nn}$ | $V_{ee}^{cl}$ | $V_{elec}^{cl}$ | $V_{xc}$ |
| | | -0.322 | -0.322 | 0.165 | 0.628 | 0.149 | -0.014 |
| | | $V_{ep}(H_1,H_2)$ | $V_{ep}(H_2,H_1)$ | $V_{np}(H_1,H_2)$ | $V_{np}(H_2,H_1)$ | $V_p^{cl}$ | |
| | -0.033 | -0.173 | -0.173 | 0.089 | 0.089 | -0.169 | |